\documentclass[12pt]{article}

\usepackage{amsmath}
\usepackage{amssymb}
\usepackage{amsfonts}
\usepackage{latexsym}
\usepackage[pdftex]{graphicx}
\usepackage{epstopdf}
\usepackage{subfigure}
\usepackage{epsfig}
\usepackage{amstext}
\usepackage[square,numbers,sort&compress]{natbib}
\usepackage[british,UKenglish,USenglish,english,american]{babel}

\def\beq{\begin{eqnarray}}
\def\eeq{\end{eqnarray}}

\begin{document}

\begin{center}

{\Large\sc The case of asymptotic supersymmetry}
\vskip 8mm


\textbf{Bruce L. S\'{a}nchez-Vega}
\footnote{
E-mail: brucesanchez@gmail.com}
\quad
\textbf{and}
\quad
\textbf{Ilya L. Shapiro}\footnote{
E-mail: shapiro@fisica.ufjf.br. On leave from Tomsk State
Pedagogical University, Tomsk, Russia.}
\vskip 4mm

Departamento de F\'{i}sica, ICE,
Universidade Federal de Juiz de Fora, 36036-330, MG, Brazil
\end{center}
\vskip 8mm

\centerline{\large\bf Abstract}
\vskip 2mm

\selectlanguage{english}
\begin{quotation}
\noindent
We start a systematic investigation of the possibility to have
supersymmetry (SUSY) to be an asymptotic state of the gauge 
theory in the high energy (UV) limit, due to the renormalization 
group running of coupling constants of the theory. The answer on 
whether this situation takes place or not, can be resolved by 
dealing with the running of the ratios between Yukawa and scalar 
couplings to the gauge coupling. The behavior of these ratios does 
not depend too much on whether gauge coupling is asymptotically 
free (AF) or not. It can be shown that the UV stable fixed point 
for the Yukawa coupling is not supersymmetric. Taking this into 
account, one can break down SUSY only in the scalar coupling 
sector. We consider two simplest examples of such breaking,
namely $N=1$ supersymmetric QED and QCD. In one of the cases one
can construct an example of SUSY being restored in the UV regime.
\vskip 6mm

\vskip 2mm

\vskip 2mm

\noindent
{\sl MSC-AMS:} \
81T60, \  
81T17, \  
81T15. \  

\end{quotation}
\vskip 4mm

 \section{Introduction}

One of the strongest theoretical motivations for SUSY is its ability to
soften the high-energy behavior of quantum field theories by reducing
the number of uncorrelated ultraviolet divergences. This is because in
supersymmetric theories there are some restrictions on both the number
of the fermionic and bosonic degrees of freedom of the theory and on
the values of the coupling constants. These restrictions provide
cancellations between fermionic and bosonic contributions to the
ultraviolet divergences, thus providing a best high-energy behavior.
Specifically, supersymmetry ensures the absence of quadratic divergences
to all orders in the loop expansion. This exciting feature of
supersymmetry offers a hope of solving the hierarchy problem \cite{gildener1976a,gildener1976b}. Roughly speaking, the solution
to this problem consists in explaining the reason for the
enormous ratio between the Planck energy scale, $\approx10^{19}$
GeV, and the energy scale $\approx 300$ GeV that characterizes the
Standard Model of the particle physics.

Even if the power-like divergences cancel, one has to deal with the
logarithmic divergences. In principle, SUSY enables one to consider
examples of a field theory with vanishing beta-functions, see, e.g., \cite{parkes1984,west1984,hamidi1984,lucha1986,ermushev1987,bohm1987}.
The most of these examples are based on the theories with extended
supersymmetry \footnote{At the one-loop level one can have finite
gauge theories without SUSY \cite{Yagu}, but the possibility of
generalization to higher loops is not certain.}.

Renormalizable quantum field theories without supersymmetry like the
Minimal Standard Model of the particle physics (MSM) suffer from the
quadratic divergence problem, although they are very successful from
the phenomenological point of view. In this type of theory, a
systematic treatment of divergent integrals in perturbation theory
was development by means of suitable renormalization schemes. Compared
to this, the gauge models with SUSY are not currently enjoying much
support from the phenomenological side, despite there are still good
chances to see SUSY at the LHC (see, e.g., \cite{SUSY_LHC}).

The world of unbroken SUSY could be perfect, but it is not a real
thing. One of the difficulties of phenomenological SUSY is that
it should be broken at low energies, at the first place this is
needed to reproduce MSM. There are different approaches to break
SUSY, e.g., dynamical \cite{Witten} and soft \cite{Dim-Geo}
breaking schemes. In both these cases the realistic SUSY imposes
some strong constrains on the parameters of the theory, such as
masses and couplings \cite{SUSY-LHC}. It is tentative to formulate
and study the following question: is it possible to reduce these
constrains such that the main advantage of SUSY, namely the
natural solution of hierarchy problem, remains valid? One of
the possibilities has been suggested long ago by Odintsov and
Shapiro \cite{odintsov1988} (see also \cite{aSUSY}), on the basis
of renormalization group running of the parameters. In this work
it was proposed to consider a rigid breaking of SUSY in the theory
with zero beta-functions and look for the situation when both SUSY
and finiteness get restored in the UV by quantum effects and
corresponding running of the couplings. It looks interesting
to generalize this result for the realistic versions of
supersymmetric Stardard Model or Grand Unification Theories
and see what would be the possible impact of the breaking of
SUSY in the couplings sector to the SUSY breaking mechanisms.

In the present work, we study the possibility that if the
restrictions on the coupling constants imposed by supersymmetry
are violated at low energies, then they can be restored in the
UV regime. To do so, we mainly use the renormalization group
method to study the behavior of the scalar coupling constants,
because the situation with Yukawa couplings is much more simple.
The result depends on the details of scalar self-interaction and
in the present work we intend to deal with the simplest possible
models, such that the situation becomes qualitatively clear. For
this reason we consider an Abelian and non-Abelian cases with the
simplest possible fields content admitting SUSY.

This paper is organized as follows. In Sect. 2 we present a brief
discussion of the running of Yukawa coupling and also formulate
the framework for dealing with scalar couplings. In Sect. 3 we
explore the behavior of the scalar coupling constants within
supersymmetric QED, assuming that these constants initially differ
from their values imposed by supersymmetry. According to what was
explained above,  in this section we simply ignore the fact that
supersymmetric QED is not AF theory. It is shown that there is an
example of asymptotic SUSY in this case, however the theory with
soft SUSY breaking meets certain physical difficulties in this case.
In Sect. 4  we analyze the same possibility in a supersymmetric
QCD and show that the situation is different in this case. Finally,
a summary and discussion of the results is given in Sect. 5.

\section{Model-indepent part: Yukawa couplings running}

Let us start by describing the way SUSY can be broken and
restored in the UV limit \cite{odintsov1988,book}. SUSY implies
rigid relations between gauge $g$, Yukawa $h$ and scalar $f$
couplings. For the sake of simplicity, consider a theory with
a single Yukawa coupling and assume that the SUSY relation
between the Yukawa and gauge couplings has the form $h=g$.
Let us now break this relation down and assume that the value
of ${\bar h}=h/g$ is arbitrary. It is well known (see, e.g.,
\cite{cheng1974}), that the general form of the renormalization
group equation (all consideration will be restricted to
one-loop order) for $h$ has the form
\beq
\frac{dh^2}{dt} = \mu\,\frac{dh^2}{d\mu}
= a_1h^4 - a_2 g^2h^2\,,
\qquad h(\mu_0) = h_0\,,
\label{Yuk-1}
\eeq
where we introduced a useful notation $t=\log (\mu/\mu_0)$
and $a_{1,2}$ are positive and otherwise model-dependent
coefficients. Taking into account the renormalization group
equation for $g(t)$,
\beq
\frac{dg^2}{dt} = \pm b^2g^4\,,
\label{Yuk-2}
\eeq
we can write the equation for the ratio in the form
\beq
\frac{d{\bar h}^2}{dt}
= g^2(t)\big[ a_1{\bar h}^4 - (a_2 \pm b^2) {\bar h}^2\big]\,.
\label{Yuk-3}
\eeq
It is easy to see that the existence of SUSY imposes a constraint
on the values of the coefficients of the last equation. One can
see that the last equation indicates a universal fixed point
${\bar h}=0$. Furthermore, we know that the value of ${\bar h}=1$
corresponding to SUSY is also a fixed point, because the theory
with SUSY is renormalizable \cite{1001,BuKu}. Therefore, the
Eq.(\ref{Yuk-3}) can be rewritten as
\beq
\frac{d{\bar h}^2}{dt}
= a_1g^2(t)\,{\bar h}^2\,\big[{\bar h}^2 - 1\big]\,.
\label{Yuk-4}
\eeq
In the case of finite $g\equiv g_0$ or asymptotically free
theory, this means that, independent on the value of $a_1$ the
SUSY fixed point is unstable in the UV limit. Any violation of
the constraint $h=g$ such that $h_0<g_0$ means that the
renormalization group trajectory for ${\bar h}(t)$ end at
the zero value in the UV fixed point. In case $h_0>g_0$ one
can observe the behavior ${\bar h}(t) \to \infty$ in the UV,
when $\,t \to \infty$, such that the SUSY is not restored
in the UV anyway. The conclusion we can draw is that the
violation of the SUSY constraint for the Yukawa coupling does
not lead to the UV restoration of SUSY. Hence in what follows
we shall always assume that the ratio $h/g=1$ is not violated
and concentrate our attention on the sector of scalar coupling
$f$, where we still have a chance to meet asymptotic SUSY.

\section{Abelian Case}

First, we consider the supersymmetric QED. The field content
consist of a Dirac field $\psi$, two complex scalar fields $\phi_+$
and $\phi_-$, a vector field $A_{\mu}$ and four-component Majorana
field $\lambda_M$. The Lagrange density has the form
\beq
\mathcal{L}_{\textrm{SQED}}
& = & -\frac{1}{4}F_{\mu\nu}F^{\mu\nu}
+ \frac{i}{2}\bar{\lambda}_{M}\gamma^{\mu}\partial_{\mu} \lambda_{M}
+ i\bar{\psi}\gamma^{\mu}D_{\mu}\psi
+ \left|D_{\mu}\phi_{+}\right|^{2}
+ \left|D_{\mu}^{\dagger}\phi_{-}\right|^{2}
\nonumber
\\
&& - \sqrt{2}g\left(\phi_{+}^{*}\bar{\lambda}_{M}\psi_{L}
+\bar{\psi_{L}}\lambda_{M}\phi_{+}
-\phi_{-}\bar{\lambda}_{M}\psi_{R}
-\bar{\psi_{R}}\lambda_{M}\phi_{-}^{*}\right)
\nonumber
\\
&& - \frac{g^{2}}{2}\left(\left|\phi_{+}\right|^{2}
-\left|\phi_{-}\right|^{2}\right)^{2} \,+\,
\textrm{(mass terms)}
\,+\, \textrm{(gauge-fixing terms)}\,,
\label{sqed}
\eeq
where $\psi_{L,\, R}\equiv\frac{1}{2}\left(1\mp\gamma_{5}\right)\psi$
and $D_{\mu}=\partial_{\mu}+igA_{\mu}$. The mass terms are not relevant
to the present analysis, and the gauge-fixing terms here are assumed
to correspond the Landau gauge. It is important that supersymmetry
imposes a relation on the 
Yukawa and scalar $\left|\phi_{\pm}\right|^{4}$ couplings, i.e., both
these couplings are proportional to the gauge coupling $g$.

Now we consider the possibility that if we break these relations between
coupling constants imposed by supersymmetry in Eq.(\ref{sqed}), they
will be recovered at high energies. In other words, will they appear
asymptotically with supersymmetry restored or not?

In order to study this possibility, we trade the term
$$
- \frac{g^{2}}{2}\left(\left|\phi_{+}\right|^{2}
-\left|\phi_{-}\right|^{2}\right)^{2}
$$
in Eq.(\ref{sqed}) for a more general expression
\beq
- \frac{f_{+}}{4}\left|\phi_{+}\right|^{4}
- \frac{f_{-}}{4}\left|\phi_{-}\right|^{4}
+ f_{\pm}\left|\phi_{+}\right|^{2}\left|\phi_{-}\right|^{2}\,.
\eeq
After this modification the theory remains renormalizable, but
it is not supersymmetric anymore. In what follows we shall consider
the one-loop approximation to the renormalization group equations
for the effective couplings $f_{+}(t)$, $f_{-}(t)$ and
$f_{\pm}(t)$ and analyze the asymptotic behavior of these effective
coupling constants. In order to perform this consideration, we divide
our analysis into three cases:
\begin{itemize}
\item Case 1: $f_{+}=f_{-}=2f_{\pm}=f$ with $f\neq2g^{2}$;
\item Case 2: $f_{+}=f$, $f_{-}=2g^{2}$ and $f_{\pm}=g^{2}$
(or equivalently
$f_{-}=f$, $f_{+}=2g^{2}$ and $f_{\pm}=g^{2}$) with $f\neq2g^{2}$;
\item Case 3: $f_{+}=f_{-}=2g^{2}$ and $f_{\pm}=f$ with $f\neq g^{2}$.
\end{itemize}
Note that, in all of the cases above, the only symmetry of the
Lagrangian in Eq.(\ref{sqed}) that has been broken is supersymmetry.
The gauge symmetry and the discrete symmetry usually called
$R-$parity are still held in the three cases.
\begin{figure}[h!]
\center
\subfigure[ref1][]{\includegraphics[width=0.19\textwidth]{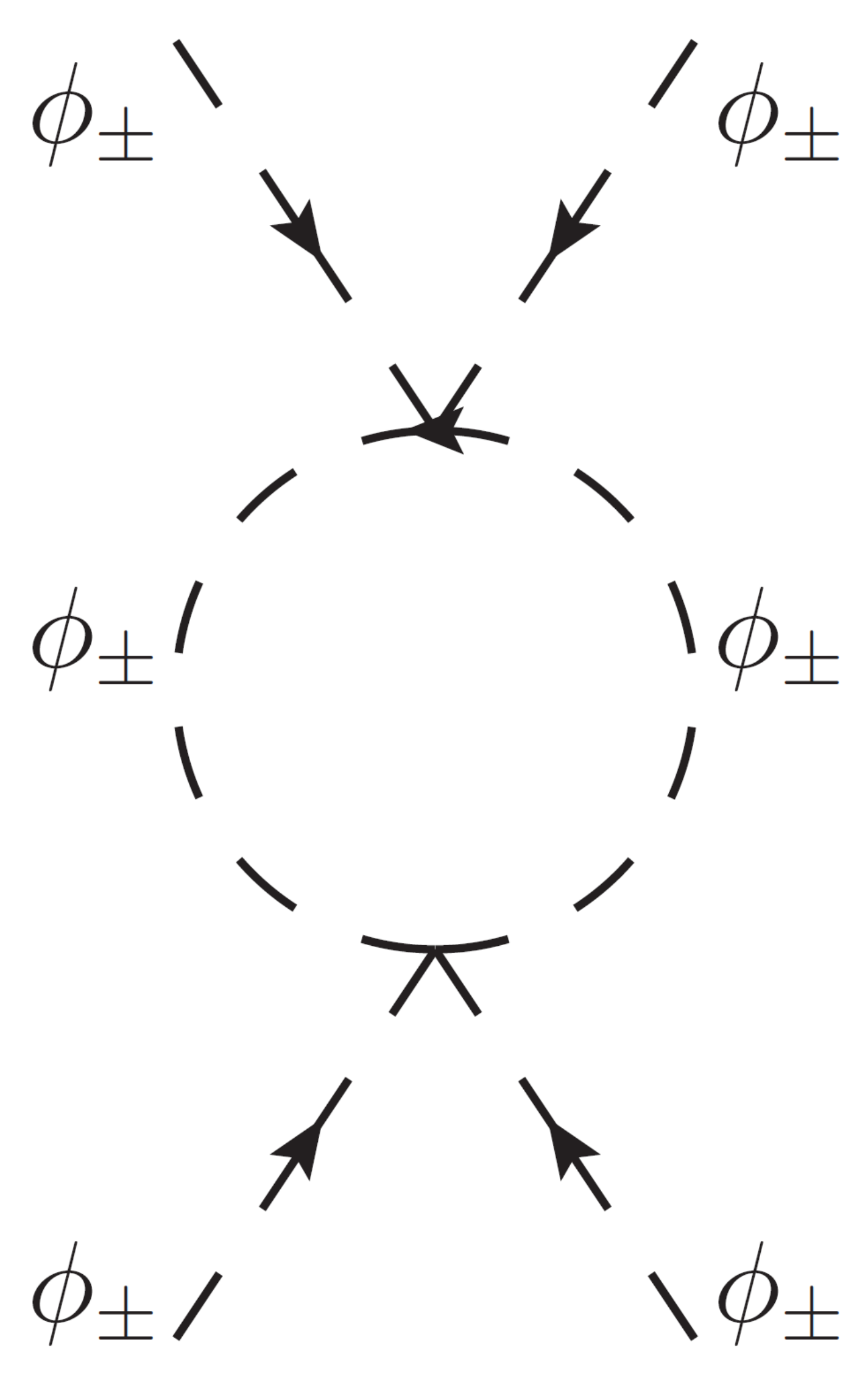}}
\qquad
\subfigure[ref2][]{\includegraphics[width=0.367\textwidth]{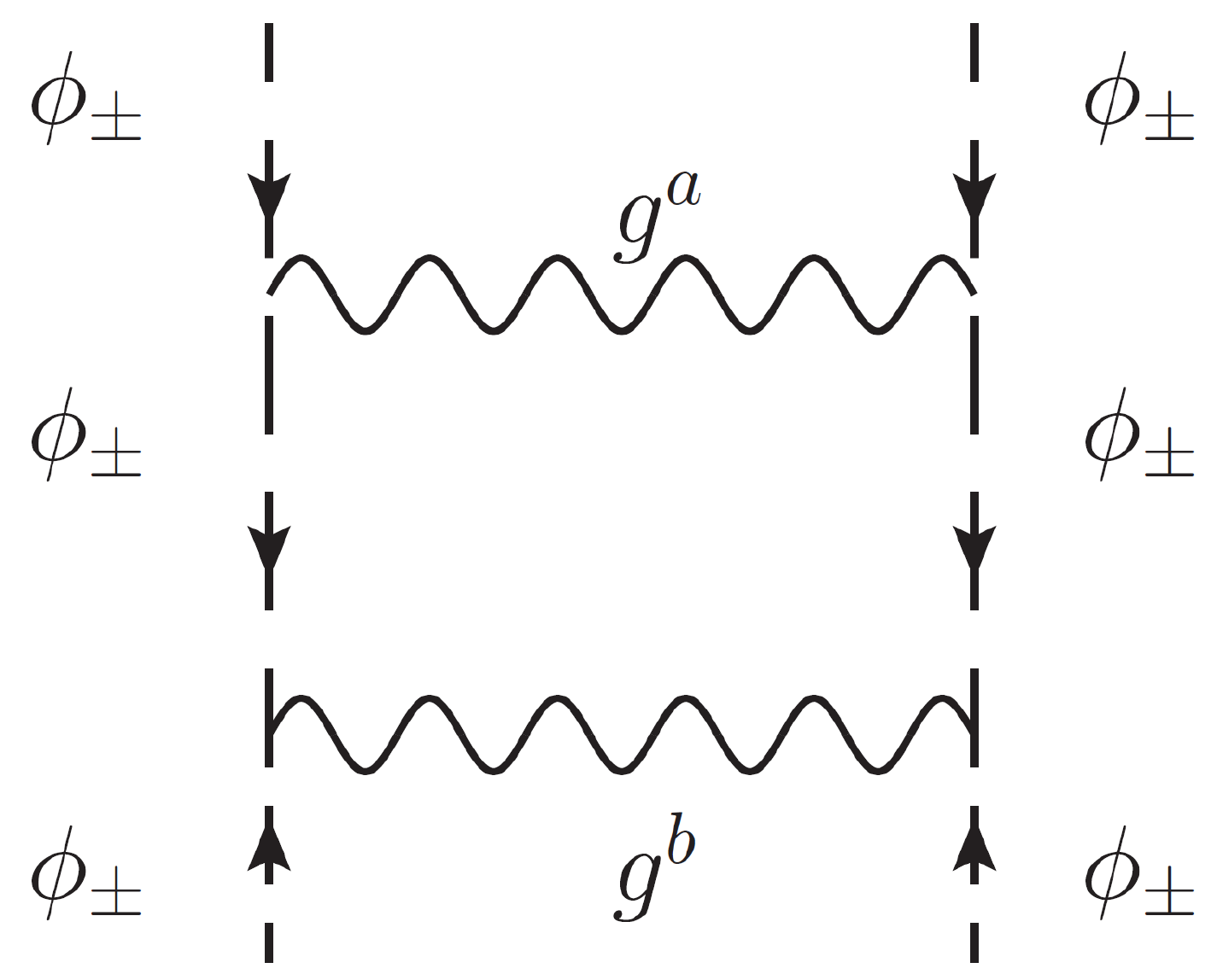}}
\qquad
\subfigure[ref3][]{\includegraphics[width=0.27\textwidth]{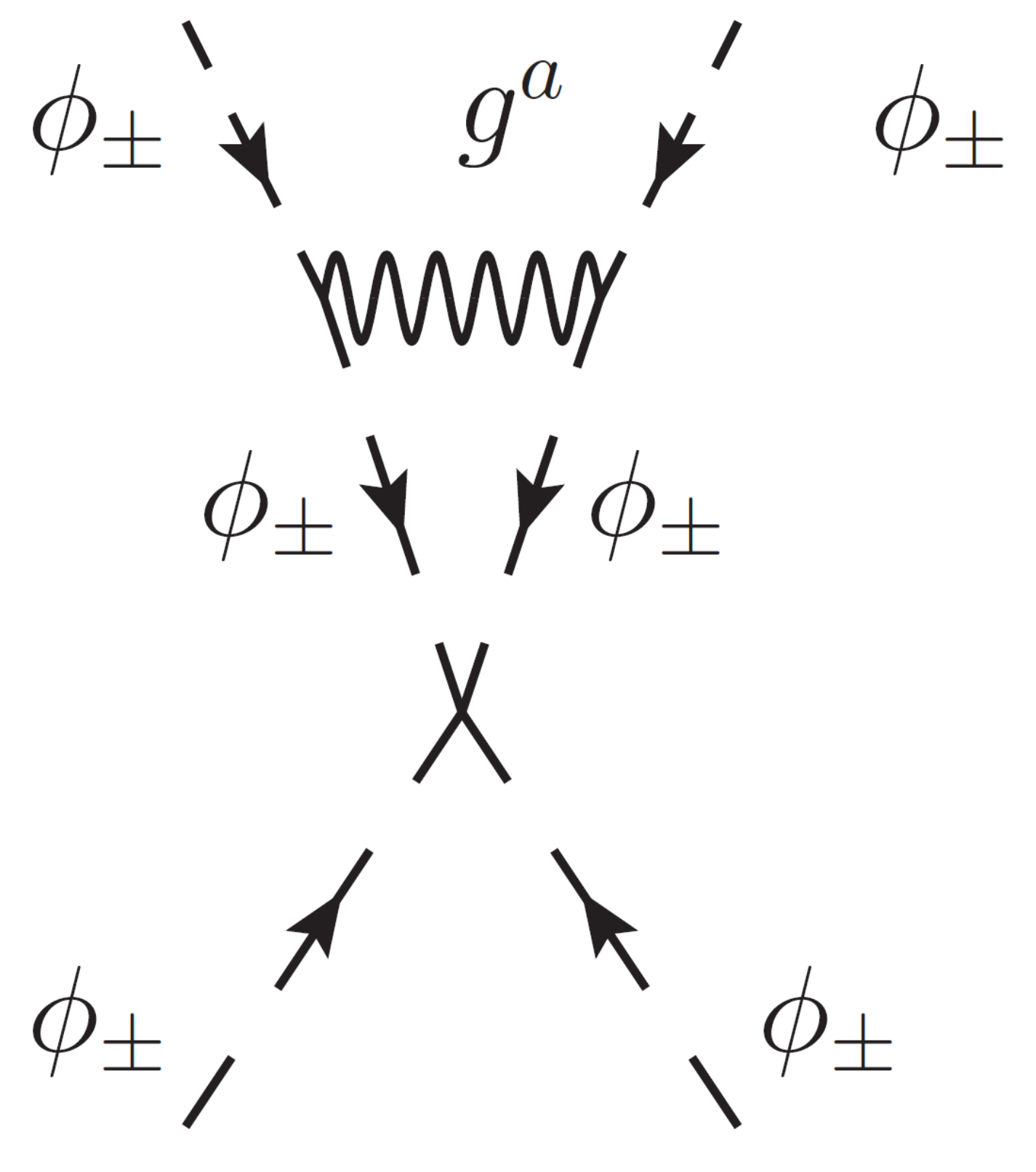}}
\newline
\subfigure[ref4][]{\includegraphics[width=0.218\textwidth]{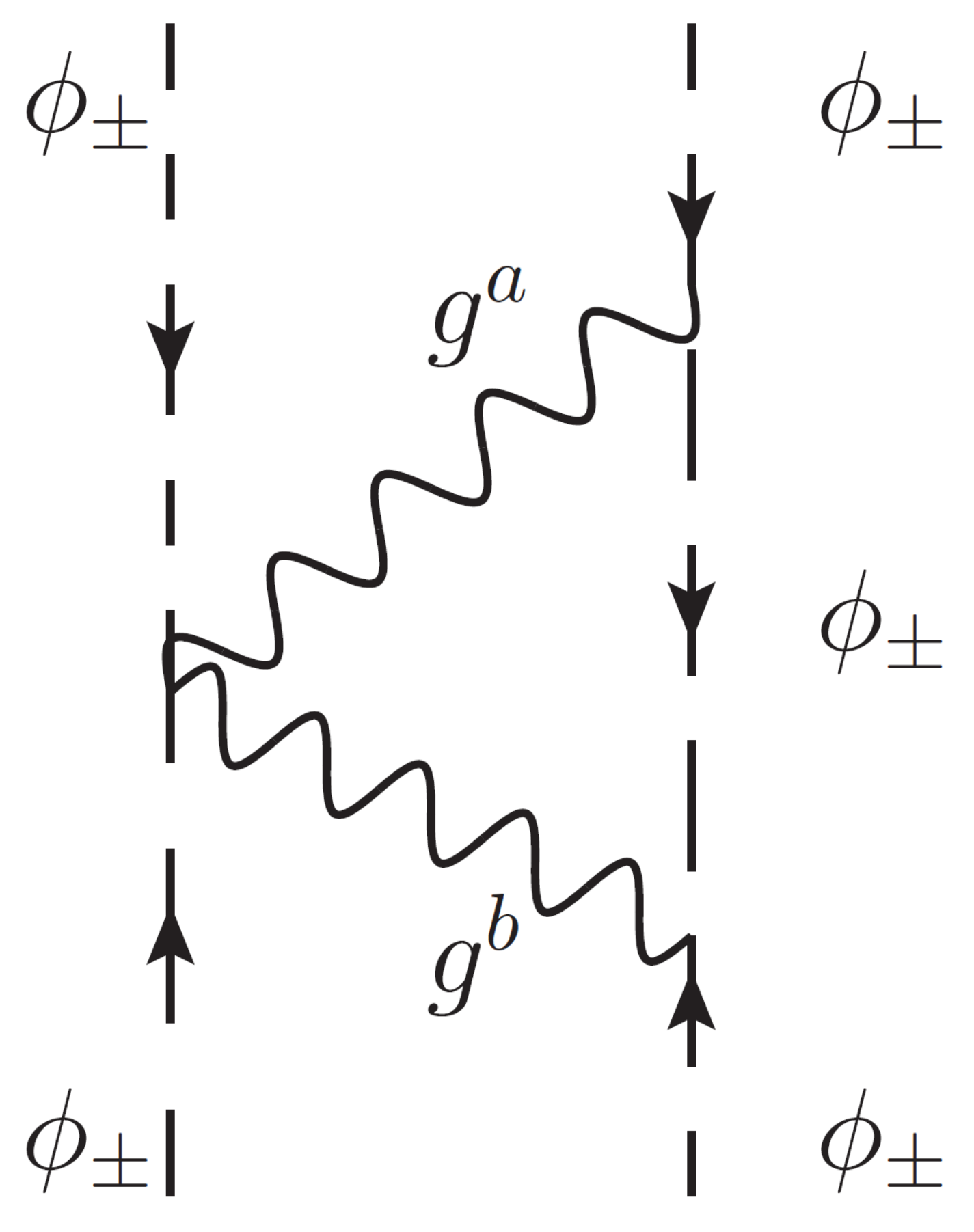}}
\qquad
\subfigure[ref5][]{\includegraphics[width=0.27 \textwidth]{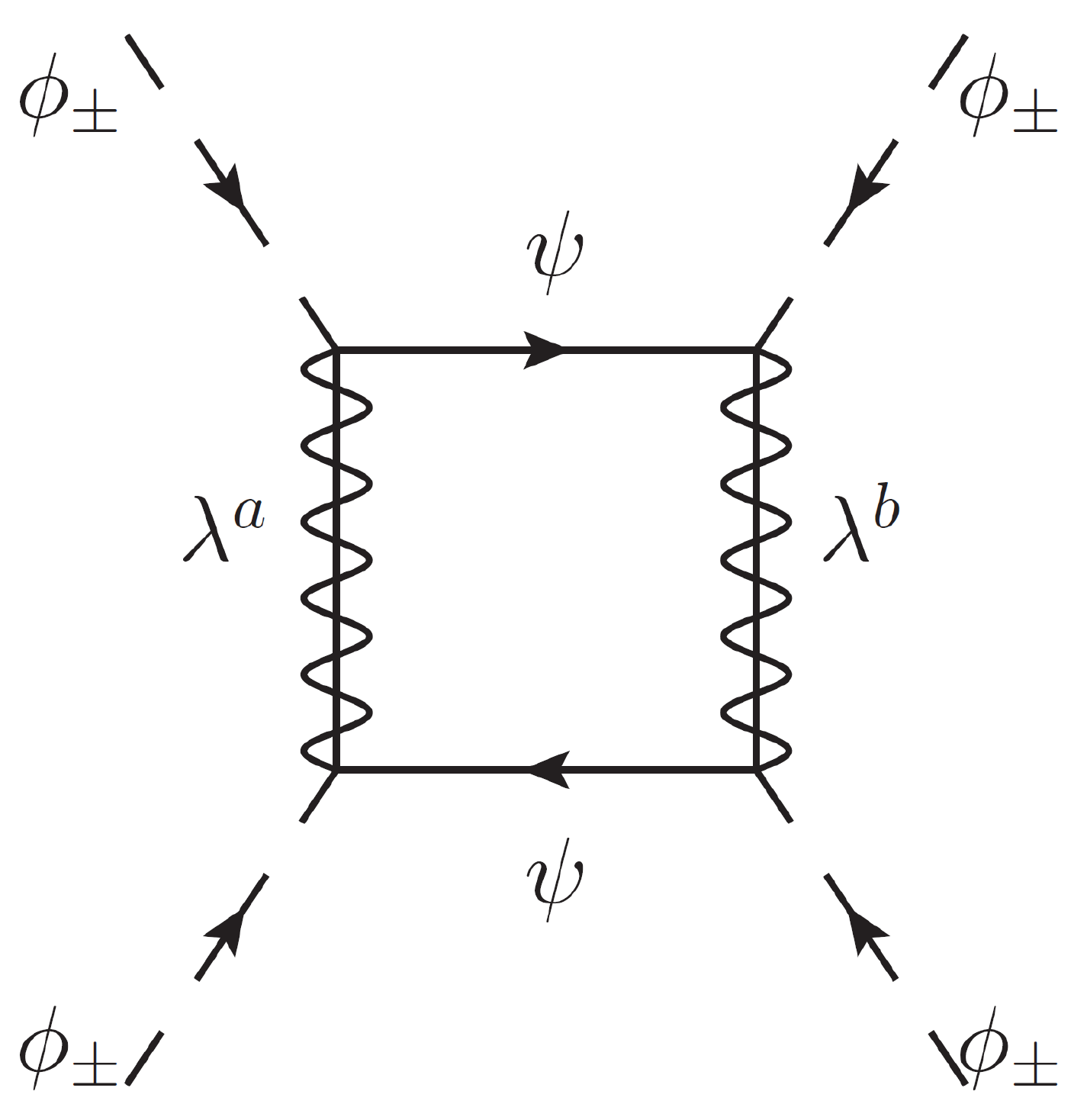}}
\qquad
\subfigure[ref6][]{\includegraphics[width=0.3 \textwidth]{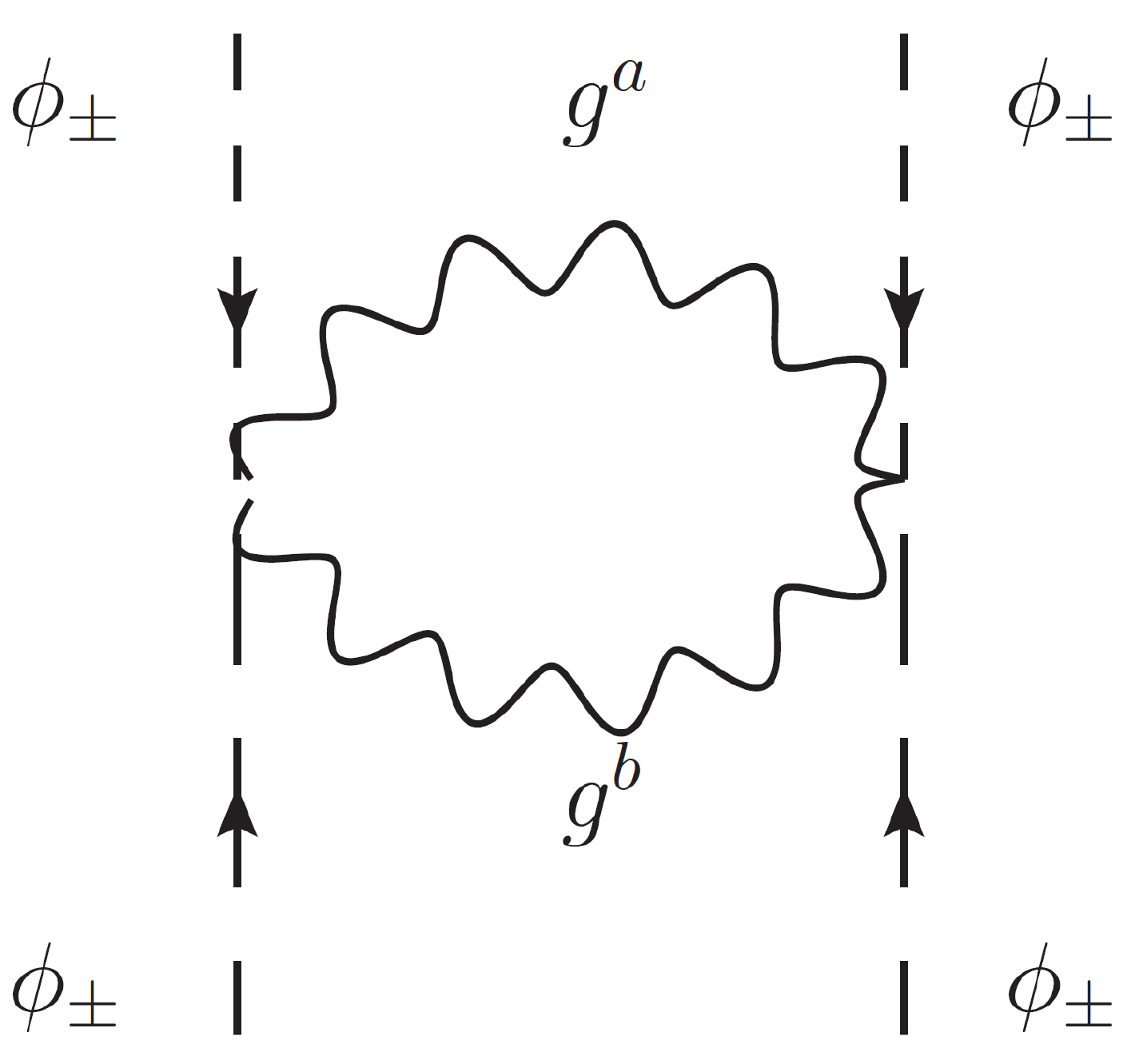}}
\begin{quotation}
\caption{One-loop diagrams contributing to the $\beta$-functions
for scalar self-couplings, discussed in the next two Sections.}\label{fig1}
\end{quotation}
\end{figure}

The diagrams needed to calculate the one-loop $\beta$-functions
in the theory with broken SUSY are presented in Fig.\ref{fig1}. We
will not discuss the details of this calculations, which are
quite standard (see, e.g., \cite{cheng1974,machacek1985}), but
go directly to the renormalization group equations of our
interest. In the Case 1, we can write down the scalar
$\beta$-function in the following form
\beq
\beta_{f}
&=&
\frac{1}{{(4\pi)^2}}\left(6f^{2}-4fg^{2}-8g^{4}\right)\,.
\label{f1}
\eeq
In order to analyze the asymptotic behavior of $f(t)$, it is
convenient to define $\bar{f}\equiv f/g^{2}$. Then we arrive
at the renormalization group equation
\beq
\frac{d\bar{f}}{dt}
\,=\,\beta_{\bar{f}}
&=&
\frac{6\,g^{2}}{(4\pi)^2}
\,\Big(\bar{f}-2\Big)\Big(\bar{f}+\frac{2}{3}\Big)\,,
\label{case1}
\end{eqnarray}
where we have used the known expression $\beta_{g}=2g^3/(4\pi)^2$.
From Eq.(\ref{case1}), we see that the $\beta_{\bar{f}}$ function
has one fixed point that correspond to the supersymmetric case ($\bar{f}_{\textrm{SUSY}}=2$),
as it should be, since the supersymmetry is not anomalous at
one-loop. At the same time, one can see that the supersymmetric
fixed point is stable in the infrared, i.e. the relation
$f/g^{2}=2$ imposed by supersymmetry at tree level is restored
when $t\rightarrow-\infty$. Concerning the UV limit\footnote{One
has to remember that we ignore the zero-charge problem here and
assume that the AF in $g(t)$ is restored by some extra contribution.}
it is easy to see that there in no asymptotic SUSY in the Case 1.

Now, let us continue with the Case 2. In this case the $\beta_{f}$
and $\beta_{\bar{f}}$ take the form
\beq
\beta_{f} & = & \frac{1}{(4\pi)^2}\left(5f^{2}-4fg^{2}-4g^{4}\right),
\nonumber
\eeq
and
\beq
\frac{d\bar{f}}{dt} \,=\,\beta_{\bar{f}}
&=& \frac{g^{2}}{(4\pi)^2}
\,5\,\left(\bar{f}-2\right)\left(\bar{f}+\frac{2}{5}\right)\,.
\label{case2}
\eeq
From Eq.(\ref{case2}), we see that in this case the supersymmetric
value, $\bar{f}_{\textrm{SUSY}}=2$ is, once again, an infrared
stable fixed point. Consequently, there is no asymptotic SUSY in the
UV limit in this case.

Finally, we find that in Case 3, the $\beta_{f}$ and
$\beta_{\bar{f}}$ functions can be written as
\beq
\beta_{f} & = & -\frac{1}{(4\pi)^2}\left(f^{2}-fg^{2}-g^{4}\right),
\nonumber
\eeq
and
\beq
\frac{d\bar{f}}{dt} \,=\,\beta_{\bar{f}}
&=& - \frac{g^{2}}{(4\pi)^2}
\left(\bar{f}-1\right)\left(\bar{f}+1\right)\,.
\label{case3}
\eeq
From Eq.(\ref{case3}) one can see that, this time, the
supersymmetric value $\bar{f}_{\textrm{SUSY}}=1$, is an
ultraviolet stable fixed point. In other words, it is an
attractive fixed point for the initial values of the
renormalization group trajectory with the initial point
satisfying the inequality $\bar{f}_0 > - 1$. In this case
$\bar{f}\rightarrow\bar{f}_{\textrm{SUSY}}$ when $t$ goes
to $\infty$.

\section{Non-Abelian Case }

As a second example, consider a supersymmetric Non-Abelian theory.
For our purposes, it is enough to take the case of such a
theory with one ``flavor'' of ``quarks'' and ``squarks'', taking
the gauge group to be $SU\left(N\right)$, with $(N\geq2)$. Here,
we are using the terms flavor, quarks and squarks, despite the
theory is reduced compared to a real QCD. Also, what we mean by
one flavor is that there is both a left-handed quark (and a
superpartner squark) in the $N-$dimensional (fundamental)
representation of the $SU(N)$ gauge group as well as a
left-handed quark in the anti-fundamental representation.
Indeed, this theory is often referred to in the literature
as supersymmetric QCD. The Lagrange density has the form
\beq
{\cal L}_{\textrm{SQCD}}
&=&
-\frac{1}{4}F_{\mu\nu}^{a}F^{\mu\nu a}
+\frac{i}{2}\bar{\lambda}_{M}^{a}\gamma^{\mu}
\left(\delta^{ac}\partial_{\mu}
- gt^{abc}A_{\mu}^{b}\right)\lambda_{M}^{c}
+i\bar{\psi}\gamma^{\mu}D_{\mu}\psi
+\left|D_{\mu}\phi_{+}\right|^{2}
+\left|D_{\mu}\phi_{-}^{*}\right|^{2}
\nonumber
\\
&& - \sqrt{2}g\left(\phi_{+}^{\dagger}\bar{\lambda}_{M}^{a}T^{a}\psi_{L}
+ \bar{\psi}_{L}T^{a}\lambda_{M}^{a}\phi_{+}
- \phi_{-}^{T}\bar{\lambda}_{M}^{a}T^{a}\psi_{R}
- \bar{\psi_{R}}\lambda_{M}^{a}T^{a}\phi_{-}^{*}\right)
\nonumber
\\
&& - \frac{1}{2}g^{2}\sum_{a=1}^{N^{2}-1}\left(\phi_{+}^{\dagger}T^{a}\phi_{+}
-\phi_{-}^{\dagger}\bar{T}^{a}\phi_{-}\right)^{2}
+ \,(\textrm{mass terms)}
\nonumber
\\
&&
+ \,\textrm{(gauge-fixing and ghost terms)}\,,
\label{sqcd}
\eeq
where $D_{\mu}=\partial_{\mu}+igT^{a}A_{\mu}^{a}$, with $T^{a}$,
$a=1,\cdots,\, N^{2}-1$, being the generators of the $SU\left(N\right)$
group in the fundamental representation and $\bar{T}^{a}$ are the
corresponding complex conjugate matrix; $t^{abc}$ are the structure
constants for $SU\left(N\right)$ in the standard form; the
hermitian conjugation and the transposition in Eq.(\ref{sqcd})
are in the $SU\left(N\right)$ internal space.

Now, we proceed by analogy with the Abelian case and trade
the scalar self-interaction term
\beq
- \frac{1}{2}g^{2} \sum_{a=1}^{N^{2}-1}
\left(\phi_{+}^{\dagger}T^{a}\phi_{+}
- \phi_{-}^{\dagger}\bar{T}^{a}\phi_{-}\right)^{2}
\nonumber
\eeq
in Eq.(\ref{sqcd}) by
\beq
-\frac{1}{4}\, f_{+}\left(\frac{N-1}{N}\right)
\left(\phi_{+}^{\dagger}\phi_{+}\right)^{2}-\frac{1}{4}\, f_{-}\left(\frac{N-1}{N}\right)
\left(\phi_{-}^{\dagger}\phi_{-}\right)^{2}\,,
\nonumber
\\
+\frac{1}{2}\, f_{1}\left(\phi_{+}^{\dagger}\phi_{-}^{*}\right)
\left(\phi_{-}^{*\dagger}\phi_{+}\right)
-\frac{1}{2N}\,f_{2}\left(\phi_{+}^{\dagger}\phi_{+}\right)
\left(\phi_{-}^{\dagger}\phi_{-}\right),
\label{potya}
\eeq
where we have used the relation
\beq
\sum\limits_{a=1}^{N^{2}-1}T_{ML}^{a}T_{NP}^{a}
=\frac{1}{2}\left(\delta_{MP}\delta_{LN}
-\frac{1}{N}\delta_{ML}\delta_{NP}\right)\,,
\nonumber
\eeq
where $M,\, P,\, L,\, N$ are the indices in the internal gauge
space. Now, for the sake of simplicity, we divide our analysis
into three cases:
\begin{itemize}
\item Case 1: $f_{+}=f_{-}=f_{1}=f_{2}=f$ with $f\neq g^{2}$;
\item Case 2: $f_{+}=f,$ $f_{-}=f_{1}=f_{2}=g^{2}$ (or equivalently $f_{-}=f$,
$f_{+}=f_{1}=f_{2}=g^{2}$) with $f\neq g^{2}$;
\item Case 3: $f_{+}=f_{-}=g^{2}$, $f_{1}\neq g^{2}$ and $f_{2}\neq g^{2}$.
\end{itemize}
In Case 1, we can write down the $\beta_{f}$ and $\beta_{\bar{f}}$
functions in the one-loop approximation as
\beq
\beta_{f} & = & \frac{1}{(4\pi)^2}\,
\left[\frac{N^{2}+4N-4}{N}f^{2}
-2\frac{N^{2}-1}{N}fg^{2}-\frac{5N^{2}
+2N-2}{N}g^{4}\right],
\nonumber
\eeq
and
\beq
\frac{d\bar{f}}{dt} \,=\,\beta_{\bar{f}}
& = & \frac{g^{2}}{(4\pi)^2}\,\frac{N^{2}+4N-4}{N}\,
\left(\bar{f}-1\right)\left(\bar{f}+\frac{5N^{2}+2N-2}{N^{2}+4N-4}\right),
\label{case1b}
\eeq
where $\bar{f}\equiv f/g^{2}$ and we also have used $\beta_{g}=-g^{3}\left(3N-1\right)/(4\pi)^2$.

For $N\geq2$, we have
\beq
\frac{N^{2}+4N-4}{N}>0
\quad \mbox{and} \quad
\frac{5N^{2}+2N-2}{N^{2}+4N-4}>0\,.
\nonumber
\eeq
Then, $\bar{f}_{\textrm{SUSY}}=1$ is an infrared-stable fixed point.

In Case 2, the $\beta_{f}$ and $\beta_{\bar{f}}$ functions
in the one-loop approximation have the following form
\beq
\beta_{f}
& = & \frac{1}{(4\pi)^2}\,\left[\frac{\left(N-1\right)
\left(N+4\right)}{N}f^{2}-2\frac{N^{2}-1}{N}fg^{2}
-\frac{5N^{2}+N-2}{N}g^{4}\right],
\nonumber
\eeq
and
\beq
\frac{d\bar{f}}{dt} \,=\,\beta_{\bar{f}}
&=& \frac{1}{(4\pi)^2}\,
\frac{\left(N-1\right)\left(N+4\right)}{N}\,
\left(\bar{f}-1\right)\left(\bar{f}
+\frac{5N^{2}+N-2}{\left(N-1\right)\left(N+4\right)}\right)\,.
\label{case2b}
\eeq
For $N\geq2$, we have
\beq
\frac{\left(N-1\right)\left(N+4\right)}{N}>0
\quad \mbox{and} \quad
\frac{5N^{2}+N-2}{\left(N-1\right)\left(N+4\right)}>0\,.
\nonumber
\eeq
Hence, once again,
$\bar{f}_{\textrm{SUSY}}=1$ is an infrared-stable fixed point.

Finally, in the Case 3 we consider simultaneously
$f_{1}\neq g^{2}$ and $f_{2}\neq g^{2}$. At the one-loop
approximation the $\beta_{f_{1}}$,
$\beta_{f_{2}}$, $\beta_{\bar{f}_{1}}$ and $\beta_{\bar{f}_{2}}$
functions can be written as
\beq
\beta_{f_{1}} & = & \frac{1}{(4\pi)^2}\,
\left[-Nf_{1}^{2}+\frac{4}{N}f_{1}f_{2}
+2\left(1-N\right)f_{1}g^{2}-\left(3N+\frac{4}{N}\right)g^{4}\right],
\\
\beta_{f_{2}} & = & \frac{1}{(4\pi)^2}
\,\left[\frac{2}{N}f_{2}^{2}+Nf_{1}^{2}
+2\left(1-N\right)f_{1}g^{2}
-\left(5N+\frac{2}{N}\right)g^{4}\right],
\eeq
and
\beq
\beta_{\bar{f}_{1}}
& = &
\frac{g^2}{(4\pi)^2}\,\left[-N\left(\bar{f}_{1}
-1\right)\left(\bar{f}_{1}-3\right)
+\frac{4}{N}\left(\bar{f}_{1}\bar{f}_{2}-1\right)\right],
\label{case3a}
\\
\beta_{\bar{f}_{2}}
& = & \frac{g^2}{(4\pi)^2}
\,\left[\frac{2}{N}\left(\bar{f}_{2}
-1\right)\left(\bar{f}_{2}+3N^{2}+1\right)
+N\left(\bar{f}_{1}-1\right)^{2}+2\left(\bar{f}_{1}
-\bar{f}_{2}\right)\right],
\label{case3b}
\eeq
where $\bar{f}_{1}\equiv f_{1}/g^{2}$ and $\bar{f}_{2}\equiv f_{2}/g^{2}$.
Using Eqs. (\ref{case3a}) and (\ref{case3b}), we can analyze the
perturbative behavior of $\beta_{\bar{f}_{1}}$ and $\beta_{\bar{f}_{2}}$
in the neighborhood of the supersymmetric fixed points, $\bar{f}_{1\,\textrm{SUSY}}=1$
and $\bar{f}_{2\,\textrm{SUSY}}=1$.

\begin{figure}[h!]
\centering
\includegraphics[width=0.7\textwidth]{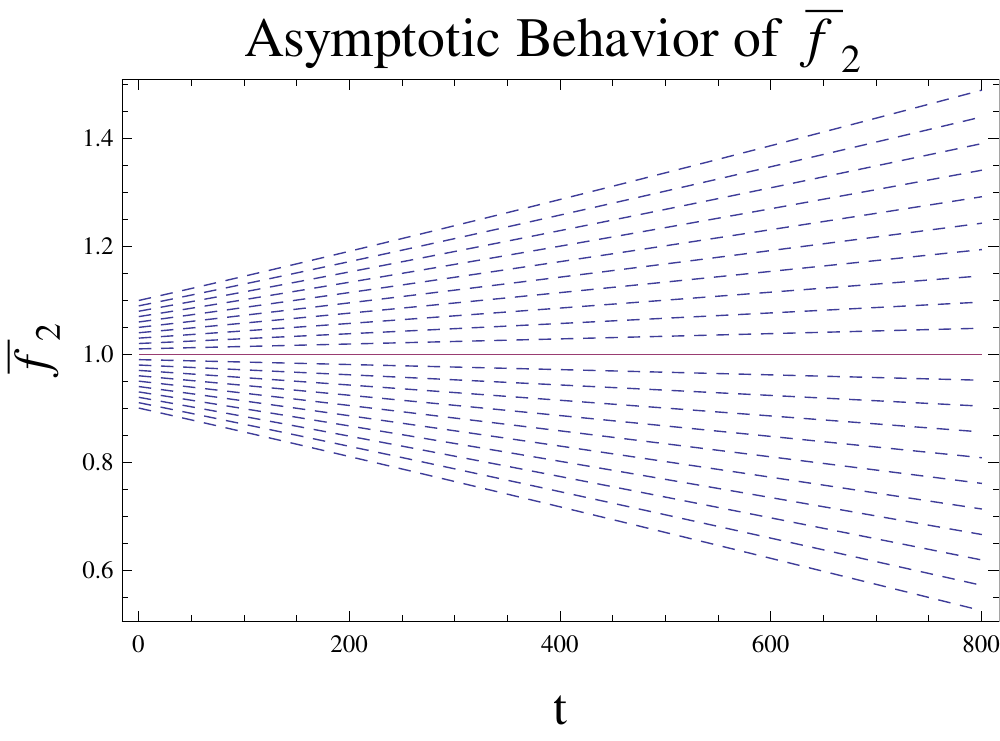}
\begin{quotation}
\caption{Behavior of $\bar{f}_{2}$   vs $t$   for supersymmetric $SU\left(3\right)$. We have used $t_{0}=0.2$, $\bar{f}_{1}\left(t_{0}\right)=1$ and $\bar{f}_{2}\left(t_{0}\right)=0.9+k\cdot\Delta$ (with $\Delta=0.01$, and $k=0,\cdots,20$) as initial conditions. The dashed lines correspond to $\bar{f}_{2}(t)$ for the different values of $\bar{f}_{2}\left(t_{0}\right)$. We see from the figure that $\bar{f}_{2\,\textrm{SUSY}}=1$ is an ultraviolet unstable fixed point.}\label{fig2}
\end{quotation}
\end{figure}

First, we consider the case that $\bar{f}_{1}=\bar{f}_{1\,\textrm{SUSY}}=1$,
but $\bar{f}_{2}\neq\bar{f}_{2\,\textrm{SUSY}}$. Note that, despite
the fact that $\bar{f}_{2}$ is not actually equal to $\bar{f}_{2\,\textrm{SUSY}}$,
we consider it to be close. Doing so, we have
\beq
\left.
\beta_{\bar{f}_{2}}\right|_{\bar{f}_{1}
=\bar{f}_{1\textrm{SUSY}}}
&=& \frac{g^{2}}{(4\pi)^2}
\,\frac{2}{N}\left(\bar{f}_{2}-1\right)
\left(\bar{f}_{2}+3N^{2}-N+1\right).
\label{betaf2}
\eeq
From Eq.(\ref{betaf2}) above, we see that $\bar{f}_{2\,\textrm{SUSY}}$
is an infrared stable fixed point. The same analysis can be done for
the case that $\bar{f}_{2}=\bar{f}_{2\,\textrm{SUSY}}$, but $\bar{f}_{1}\neq\bar{f}_{1\,\textrm{SUSY}}$
. In this case, we have
\beq
\left.\beta_{\bar{f}_{1}}\right|_{\bar{f}_{2}
= \bar{f}_{2\textrm{SUSY}}}
&=&
-\frac{g^{2}}{(4\pi)^2}\, N \left(\bar{f}_{1}-1\right)
\Big[\bar{f}_{1} - \Big(3 + \frac{4}{N^2}\Big)\Big]\,.
\label{betaf1}
\eeq
As in the previous case, from Eq.(\ref{betaf1}), one can see that
$\bar{f}_{1\,\textrm{SUSY}}$
is an infrared stable fixed point. In order to illustrate the behavior
of $\bar{f}_{2}$ in function of $t$, we have numerically integrated
the Eqs.(\ref{case3a}) and ($\ref{case3b}$) for the particular case
of supersymmetric $SU\left(3\right)$. We have used the initial
conditions
\ $g\left(t_{0}\right)=0.2$,
\ $\bar{f}_{1}\left(t_{0}\right)=1$ and $\bar{f}_{2}\left(t_{0}\right)=0.9+k\cdot\Delta$,
where $\Delta=0.01$ and $k=0,\cdots,20$.
The plots illustrating the running of $\bar{f}_{2}$ are shown
in the Fig.\ref{fig2}.

\section{Conclusions}

The theories with supersymmetric particle/field content, but with
a non-supersymmetric choice of coupling constants are potentially
interesting, especially because SUSY may get restored
asymptotically, in the UV limit \cite{odintsov1988}. If this
asymptotic behavior really takes place, one may hope to solve the
hierarchy problem with less restrictions on the parameters of the
theory, which maybe a welcome feature from the phenomenological
viewpoint. In order to address this possibility, we have considered
a one-loop running in the models with violated SUSY. The fate of
the violation in the Yukawa sector can be traced back in the
general form, and we conclude that the SUSY is always an IR
fixed point, independent on the details of the theory. Then our
main attention has been paid to the situation with a violation in
the sector of scalar self-interaction couplings.

To summarize our results, we have shown that in both Abelian and
non-Abelian cases, the function $\beta_{\bar{f}}$, at one-loop
approximation, has the following form:
\beq
\beta_{\bar{f}}
& = & c_1\,
\left(\bar{f}-\bar{f}_{\textrm{SUSY}}\right)
\left( \bar{f}-\bar{f}_1 \right)\,,
\label{betafinal}
\end{eqnarray}
where $c_{1}$ is the model-dependent coefficient and the value
of $\bar{f}_1$ corresponds to a non-supersymmetric version of the
gauge theory.

We have considered several particular ways to violate SUSY
in both Abelian and non-Abelian models. In most cases we have
found that the two fixed point satisfy the inequality
$\bar{f}_{\textrm{SUSY}}>\bar{f}_1$. For $c_1 > 0$
this means that the SUSY is an IR-stable fixed point, and not
an UV-stable. Obviously, the IR running based on the Minimal
Subtraction renormalization scheme is not a clear issue, because
the sparticle are supposed to be massive and, therefore, the
running (if any) should be strongly modified by decoupling
of massive degrees of freedom.

Only in one particular case, with Abelian gauge group, we
have found the UV-stable asymptotic SUSY, due to the negative
sign of the $c_1$ coefficient. Let us note that the negative
sign here does not lead to problems with stability, because
other $\beta$-functions remain positive. The real advantage
of dealing with Abelian model is its simplicity, which gives
an opportunity to better control the fixed points for the
ratio between scalar and gauge couplings. The fact that the
theory under consideration is Abelian does not look too
important for the distribution of zeroes of $\beta_{\bar{f}}$,
so we have simply ignored the positive sign of the
$\beta$-function for the gauge coupling. Of course, it
would be interesting to extend the present consideration
to the more realistic cases of the gauge models with
possible asymptotic SUSY.

In a similar situation (Case 3) in the supersymmetric QCD we
have found $c_1<0$, but unfortunately
with $\bar{f}_{\textrm{SUSY}}<\bar{f}_1$. As a consequence, $\bar{f}_{\textrm{SUSY}}$ is not
an UV stable fixed point in this case. The particular examples
considered here are not sufficient to see whether the consistent
solution with UV asymptotic SUSY is possible or not, but is
worthwhile to study this subject further.

All in all, it remains unclear whether the asymptotic SUSY
can be achieved within some version of realistic supersymmetric
gauge theory, including non-minimal Standard Model. Our
consideration shows that this possibility can not be ruled
out and therefore some further work in this direction looks
interesting to do. In case of a positive output, the
corresponding gauge theory would be a promising model to
study particle phenomenology.

\section*{Acknowledgments}

B. L. S\'{a}nchez--Vega would like to thank J. A. Helay\"{e}l-Neto
for valuable discussions and the Departamento de F\'{\i}sica
of UFJF for kind hospitality. His work was supported by CAPES
through the PNPD program. I.Sh. is thankful to CNPq, FAPEMIG
and ICTP for partial support of his work.


\end{document}